\documentclass[one column,showpacs,floatfix,11pt,nofootinbib]{revtex4-1}
\usepackage{amssymb,amsmath}
\usepackage{graphicx,float}
\usepackage{indentfirst}
\newcommand{\be}{\begin{equation}}
\newcommand{\ee}{\end{equation}}

\newcommand{\ba}{\begin{eqnarray}}
\newcommand{\ea}{\end{eqnarray}}

\begin{document}
\title{Black Strings from Minimal Geometric Deformation in a Variable Tension Brane-World}
\author{R. Casadio}
\email{casadio@bo.infn.it}
\affiliation{Dipartimento di Fisica e Astronomia, Universit\`a di Bologna, 
via Irnerio 46, 40126 Bologna, Italy
\\
Istituto Nazionale di Fisica Nucleare, 
Sezione di Bologna, viale B.~Pichat 6, 40127 Bologna, Italy}
\author{J. Ovalle}
\email{jovalle@usb.ve}
\affiliation{Departamento de Fisica, Universidad Sim\'on Bol\'ivar, Apartado 89000, Caracas 1080A, Venezuela}
\author{Rold\~ao da Rocha}
\email{roldao.rocha@ufabc.edu.br}
\affiliation{Centro de Matem\'atica, Computa\c c\~ao e Cogni\c c\~ao, Universidade Federal do ABC (UFABC) 09210-580, Santo Andr\'e, SP, Brazil.}
\begin{abstract}
We study brane-world models with variable brane tension and compute corrections 
to the horizon of a black string along the extra dimension.
The four-dimensional geometry of the black string on the brane is obtained by means of
the minimal geometric deformation approach, and the bulk corrections are then encoded
in additional terms involving the covariant derivatives of the variable brane tension.
Our investigation shows that the variable brane tension strongly affects the shape and evolution
of the black string horizon along the extra dimension, at least in a near-brane expansion.
In particular, we apply our general analysis to a model motivated by the E\"otv\"os branes,
where the variable brane tension is related to the Friedmann-Robertson-Walker brane-world
cosmology.
We show that for some stages in the evolution of the universe, the black string warped horizon
collapses to a point and the black string has correspondingly finite extent along the extra dimension.
Furthermore, we show that in the minimal geometric deformation of a black hole on the variable
tension brane, the black string has a throat along the extra dimension, whose area tends to zero
as time goes to infinity. 
\end{abstract}
\maketitle
\flushbottom
%
%
%
\section{Introduction}
Brane-world (BW) models~\cite{RS} represent a distinct branch of contemporary high-energy
physics, inspired and supported by string theory.
Indeed, these models are a straightforward five-dimensional phenomenological realization of the
Ho$\check{\rm r}$ava-Witten supergravity solutions~\cite{HW}, when the hidden brane is
moved to infinity, and the moduli effects from the further compact extra dimensions may
be neglected (for a review, see e.g.~Refs.~\cite{maartens}).
Beside several precursory papers about the BW universe~\cite{PREV},
there is nowadays a great variety of alternative models which have been developed in order
to address specific aspects.
For example, some studies focus on developing the dimensional reduction stemming
from a strongly curved extra-dimension, others provide a realization of the AdS/CFT
correspondence to lowest order, and further incorporate the self-gravity of the brane,
namely the brane tension $\sigma$.
\par
Among all BW models, those containing a brane endowed with variable tension,
that is $\sigma = \sigma(x^\mu)$, where $x^\mu$ are coordinates on our four-dimensional
brane, are a noteworthy attempt to ascertain signatures coming from
high-energy physics beyond the Standard Model~\cite{VACARU}.
In fact, since the temperature has changed dramatically along the cosmological evolution,
a variable tension BW scenario is indeed a natural candidate to describe our
physical universe.
The fully covariant dynamics of variable tension branes was established in Ref.~\cite{GERGELY2008},
and further explored in Ref.~\cite{GERGELY2009}.
Moreover, a variable tension was also implemented in BW model consisting of two
branes~\cite{PRD}, and the cosmological evolution was investigated in a 
particular model in which the brane tension has an exponential dependence on
the scale factor~\cite{EPJC}.
\par
In this work, we consider black strings in BW models with variable brane tension,
whose four-dimensional geometry is determined according to the principle of minimal
geometric deformation (MGD)~\cite{jovalle2009}.
In particular, we shall focus on the shape and evolution of the black string time-dependent horizon,
and show that, in this setup, the black string warped horizon is drastically affected by the variable
tension, since additional bulk terms are induced by the variable brane tension.
For this purpose, we shall employ a Taylor expansion in the extra dimension about
a black hole metric on the brane, in such a away that the corrections to the area of the five-dimensional
black string horizon are singled out.
A complete numerical analysis will then be displayed for a model physically motivated by
E\"otv\"os law.
\par
Approximative methods involving expansions of the metric have been extensively used
in order to investigate black strings in the BW.
For instance, previous constructions of black holes and black strings in vacuum plane wave
space-times employed the so called method of matched asymptotic expansions~\cite{LeWitt:2009qx}.
Another construction was given in Ref.~\cite{Haddad:2012ss}, based on a derivative expansion
method along the direction of the black string, which provided a solution up to second order
in derivatives and which showed, in particular, that the black string must shrink to zero size
at the horizon of some black brane.
In Ref.~\cite{Haddad:2012ss}, the usual black string metric thus appears as the leading order
solution in a given expansion that contains corrections to the black string metric order by order
in derivatives.
Other approximative methods, including Taylor and Fourier metric expansions,
have been employed to provide black string profiles~\cite{empa,bha}.
Such expansions in the context of general relativity (GR) and black strings are thus commonly
used, as, schematically, the degrees of freedom of GR are split into long and short wavelength
components.
Also, approximations usually concern the far-zone and near-zone metric for the black string,
the two regimes being defined where the radial coordinate is respectively much larger
or much smaller than twice the black hole mass~\cite{Emparan:2009at}.
\par
Regarding the time evolution of black strings, some seminal results showed that their
warped horizon cannot pinch in a finite horizon time~\cite{Horowitz:2001cz,kol}.
This means that a black hole cannot be the end-state of their Hawking decay and, instead,
the existence of a stable string phase should serve as an end-state.
The argument that shows the pinching takes an infinite time is based on assuming
the increasing area theorem for an event horizon and applying it to an area element
at the throat, namely, the inward collapsing region of the event horizon.
In addition, numerical analyses also support the conclusion that the warped horizon 
cannot pinch in finite time~\cite{kol}. 
\par
Our approach here is much more general, in that it provides the bulk metric near the brane,
using Gaussian normal coordinates in order to extend into the bulk a BW black hole metric
obtained from the MGD of the Schwarzschild metric.
Computing the bulk metric when the radial coordinate equals the horizon radius of
the brane black hole, will thus provide a description of the black string warped horizon,
just as a particular case.
Let us stress that the bulk shape of the horizon has only been investigated in very particular
cases~\cite{maartens,casadio1}, and, recently, the Schwarzschild black string
was studied in this context, e.g.~for a brane with variable tension~\cite{bs1}.
Some realistic generalizations regarding a post-Newtonian parameter on the
Casadio-Fabbri-Mazzacurati black string~\cite{cfabbri,Bazeia:2013bqa},
and the  black string in a Friedmann-Robertson-Walker 
BW~\cite{daRocha:2013ki} also represent interesting applications.   
\par
In the next Section, we shall recall the general expansion of the bulk metric elements
in the Gaussian normal coordinate perpendicular to the brane and, in Section~\ref{MGD},
we shall review the MGD approach to generate BW solutions starting from GR solutions.
In Section~\ref{BS}, we shall then apply the latter method in order to obtain all the terms that
determine the expansion of the bulk metric up to forth order, starting from the Schwarzschild
metric on a brane with variable tension.
We shall finally consider the particular case of E\"otv\"os fluid branes endowed with a tension
derived from a de~Sitter-like cosmological scale factor, and investigate the role that the
variable brane tension plays in this context.
The black string warped horizon along the extra dimension will be analysed, and we shall
show that our construction based on the MGD induces a throat on the black string,
and prevents the black string warped horizon to pinch in finite horizon time. 
\section{Bulk metric in Gaussian coordinates}
\label{bulkmetric}
In this Section, the general formalism of Ref.~\cite{maartens} is briefly introduced
and reviewed.
Hereon, $\{\theta_\mu\}$, with {$\mu = 0,1,2,3$} [and $\{\theta_A\}$, with {$A=0,1,2,3,5$}]
denotes a basis for the cotangent space $T^\ast_x\mathcal{M}$ at a point $x$ on a 3-brane
$\mathcal{M}$, embedded in the five-dimensional bulk.
Furthermore, $\{e_A\}$ is its dual basis and $\theta^A = dx^A$, when a coordinate chart is chosen.
Let $n = n_A\,\theta^A$ be a time-like covector orthogonal to $T^\ast_x\mathcal{M}$
and $y$ the associated Gaussian coordinate, which parameterizes geodesics starting
from the brane and moving into the bulk.
In particular, $n_A\,dx^A = dy$ on the hypersurface defined by $y=0$.
The five-dimensional coordinate vector field $u = x^A\,e_A$ thus splits into components
parallel and orthogonal to the brane, and can be written as  $u = x^\mu \,e_\mu + y\,e_5$
or $u^A= (x^{\mu},y)$.
The brane metric $g_{\mu\nu}$ and the corresponding components of the bulk metric
$\check{g}_{\mu\nu}$ are in general related by
$\check{g}_{\mu\nu} = g_{\mu\nu} + n_\mu\, n_\nu$~\cite{maartens}.
Since with our choice of $n$ we have $g_{55} = 1$ and $g_{\mu 5} = 0$, the five-dimensional
bulk metric $\check{g}_{AB}\,dx^A\, dx^B = g_{\mu\nu}(x^\alpha,y)\,dx^\mu\,dx^\nu + dy^2$,
and one can effectively use $A,B = 0,1,2,3$.
\par
There is a well known relation between the effective four-dimensional cosmological constant
$\Lambda_4$ on the brane, the bulk cosmological constant $\Lambda_5$, and the brane tension
$\sigma$, given by the fine-tuning relations~\cite{maartens}
\be
\kappa^2_{4}=\frac{1}{6}\sigma\kappa^4_5\,,
\qquad\qquad\quad
\Lambda_4=\frac{\kappa_5^{2}}{2}\left(\Lambda_{5}+\frac{1}{6}\kappa_5^{2}\sigma^{2}\right)
\ ,
\ee
where $\kappa_5$ ($\kappa_4$) denotes the five-dimensional (four-dimensional)
gravitational coupling. 
The extrinsic curvature of the brane at $y=0$ is given by
$K_{\mu\nu} = \frac{1}{2}\frac{\partial}{\partial y} g_{\mu\nu}$ in Gaussian
normal coordinates, and the junction conditions thus imply that
\be
\label{curv}
K_{\mu\nu}
=
-\frac{\kappa_5^2}{2} \left[T_{\mu\nu}+ \frac{1}{3}
\left(\sigma-T\right)g_{\mu\nu} \right]
\ .
\ee
The symmetric and trace-free components of the bulk Weyl tensor
$
 C_{\mu\nu\sigma\rho}$ are respectively
given by
${\cal E}_{\mu\nu} = C_{\mu\nu\sigma\rho} \,n^\sigma\, n^\rho$ 
and
${\cal B}_{\mu\nu\alpha} = g_\mu^{\;\rho}\, g_\nu^{\;\sigma} \,C_{\rho\sigma\alpha\beta}\,n^\beta$.
\par
The effective four-dimensional field equations are complemented by a set of equations
obtained from the five-dimensional Einstein and Bianchi equations in
Refs.~\cite{GCGR,Gergely:2003pn,maartens} for constant $\sigma$,
and in Ref.~\cite{GERGELY2008} for a brane with variable tension,
all approaches being completely consistent.
In particular, the bulk metric near the brane can be expressed as the Taylor expansion
in the Gaussian coordinate $y$ as
\be
g_{\mu\nu}(x^\alpha,y)
=
\sum_p\,
g_{\mu\nu}^{(p)}(x^\alpha)\,\frac{|y|^p}{p!}
\ .
\ee
By denoting $g_{\mu\nu}\equiv g_{\mu\nu}^{(0)}(x^\alpha)$,
and $R_{\mu\nu\sigma\rho}\equiv R_{\mu\nu\sigma\rho}(x^\mu,0)$
the components of the bulk Riemann tensor ($R_{\mu\nu}$ and $R$
are obviously the associated Ricci tensor and scalar curvature) likewise
computed on the brane,
one then finds the above expansion up to order $p=4$ is given by
\ba
g_{\mu\nu}(x^\alpha,y)
&=&
g_{\mu\nu}
-\kappa_5^2\left[
T_{\mu\nu}+\frac{1}{3}(\sigma-T)g_{\mu\nu}\right]\,|y| 
\nonumber
\\
&&
+\left[\frac{1}{2}\kappa_5^4\left(
T_{\mu\alpha}T^\alpha{}_\nu +\frac{2}{3} (\sigma-T)T_{\mu\nu}
\right) -2{\cal E}_{\mu\nu} +\frac{1}{3}\left( \frac{1}{6}
\kappa_5^4(\sigma-T)^2-\Lambda_5
\right)g_{\mu\nu}\right]\, \frac{y^2}{2!}
\nonumber
\\
&&
+\left.\Bigg[2K_{\mu\beta}K^{\beta}_{\;\,\alpha}K^{\alpha}_{\;\,\nu} - 
{\cal E}_{(\mu\vert\alpha}K^{\alpha}_{\;\,\vert\nu)}
-\nabla^\rho{\cal B}_{\rho(\mu\nu)} + \frac{1}{6}
\Lambda_5g_{\mu\nu}K+K^{\alpha\beta}R_{\mu\alpha\nu\beta}-K{\cal E}_{\mu\nu}
\right.
\nonumber
\\
&&
\left.
\quad
\qquad\qquad+3K^\alpha{}_{(\mu}{\cal
E}_{\nu)\alpha}+K_{\mu\alpha}K_{\nu\beta}K^{\alpha\beta}
-K^2K_{\mu\nu}\Bigg]\;\frac{|y|^3}{3!}
\right.
\nonumber
\\
&&
+\left.\Bigg[\frac{\Lambda_5}{6}\left(R-\frac{\Lambda_5}{3} + K^2\right)g_{\mu\nu} + \left(\frac{K^2}{3}
-\Lambda_5\right)K_{\mu\alpha}K^{\alpha}_{\;\,\nu} + (R-\Lambda_5 + 2K^2){\cal E}_{\mu\nu}\right.
\nonumber
\\
&&
+
\left. \left(K^{\alpha}_{\;\,\tau}K^{\tau\beta} + {\cal E}^{\alpha\beta}
+KK^{\alpha\beta}\right)\,R_{\mu\alpha\nu\beta} + K^2\,K\,K_{\mu\nu}- \frac{1}{6}\Lambda_5R_{\mu\nu}
 + 2 K_{\mu\beta}K^{\beta}_{\;\,\rho}K^\rho_{\;\,\alpha}K^\alpha_{\;\,\nu} \right.
\nonumber
\\
&&
+\left.
 {\cal E}_{\mu\alpha}\left(\frac{1}{2}KK^\alpha_{\;\,\nu}-3K^\alpha_{\;\,\sigma}K^{\sigma}_{\;\,\nu} \right)-\frac{13}{2}K_{\mu\beta}{\cal E}^\beta_{\;\,\alpha}K^{\alpha}_{\;\,\nu} - 4K^{\alpha\beta}R_{\mu\nu\gamma\alpha}K^{\gamma}_{\;\beta} - K_{\mu\alpha}K_{\nu\beta}{\cal E}^{\alpha\beta}
\right.
\nonumber
\\
&&
\left.
+\frac{7}{2}KK^\alpha_{\;\,\mu} 
{\cal E}_{\nu\alpha}
- \frac{7}{6}K^{\sigma\beta}K^{\;\,\alpha}_{\mu}R_{\nu\sigma\alpha\beta}\Bigg]\,\frac{y^4}{4!}
+\cdots
\right.
\ ,
\label{tay}
\ea
where $B\equiv B^\mu{}_\mu $ and  $B^2 \equiv B_{\alpha\beta}\,B^{\alpha\beta}$, for any rank-2
tensor $B$. 
This expansion was analysed in Refs.~\cite{maartens,casadio1} only up to the second order,
but this is not sufficient to determine the additional terms arising from the brane variable tension.
Alternative approaches do not take into account the $\mathbb{Z}_2$ symmetry ~\cite{Jennings:2004wz}.
\par
The additional terms coming from the variable brane tension shall indeed be shown to play
an essential role for the subsequent analysis of the black string behaviour along the extra dimension.
In particular, terms in the expansion~\eqref{tay} involving derivatives of the variable brane tension 
at order $|y|^3$ are given by~\cite{bs1}
\ba
g_{\mu\nu}^{(3)\,{}^{\rm variable}}
=
\frac{2}{3}\kappa_5^2\left[\nabla_{(\nu}\nabla_{\mu)}\sigma
-
g_{\mu\nu}\,\Box\sigma\right]
\ ,
\label{addtruey3}
\ea
and, at order $y^4$, by
\ba
g_{\mu\nu}^{(4)\,{}^{\rm variable}}
&=&
- \frac{\kappa_5^2}{3}\left[\Box(\Box\sigma)g_{\mu\nu}-\nabla_{(\nu}\nabla_{\mu)}(\Box\sigma)\right]
+\left(\frac{1}{3}\kappa_5^2+2 K\right)
\left\{(\Box\sigma){\cal E}_{(\mu\nu)} - \nabla^\alpha\left[(\nabla_{(\mu}\sigma)\, {\cal E}_{\nu)\alpha}\right]\right\}
\nonumber
\\
&&
+ \frac{\kappa_5^2}{3}
\left\{(\Box\sigma)R_{\mu\nu}-\nabla^\alpha\left[(\nabla_{(\mu\vert}\sigma)\,R_{\alpha\vert\nu)}\right]\right\}
-2 K^{\tau\beta}
\left\{(\Box\sigma)R_{(\mu\vert\tau\vert\nu)\beta}
- \nabla^\alpha\left[(\nabla_{(\mu\vert}\sigma)\, R_{\alpha\tau\vert\nu)\beta}\right]\right\}
\nonumber
\\
&&
+\frac{\kappa_5^2}{3}
\left\{(\Box\sigma)\left(K_{(\mu\vert\rho}K_{\vert\nu)\beta} K^{\rho\beta} - K^2\,K_{\mu\nu}\right)
- \nabla^\alpha
\left[(\nabla_{(\mu\vert}\sigma)\left(K_{\alpha\sigma}K^{\;\,\sigma}_{\vert\nu)} - K\,K_{\alpha\vert\nu)}\right)
\right]
\right\}
\nonumber
\\
&&
+6 \left\{
(\Box\sigma)K_{(\mu\vert\tau}{\cal E}_{\vert\nu)}^\sigma
- \kappa_5^2\nabla^\alpha\left[
(\nabla_{(\mu}\sigma)\, {\cal E}_{\nu)\alpha}\right]
\right\}
+\left(2\,K^2-\frac{1}{3}\Lambda_5 \right)
\left[(\Box\sigma)g_{\mu\nu}-\nabla_{(\nu}\nabla_{\mu)}\sigma\right]
\nonumber
\\
&&
+2\left(K + \frac{7}{3}\kappa_5^2\right)
\left\{(\Box\sigma)K\,K_{\mu\nu}
-\nabla^\alpha\left[(\nabla_{(\mu\vert}\sigma)\, K\,K_{\alpha\vert\nu)}\right]
\right\}
\ .
\label{addtruey4}
\ea
In the following, we shall just consider a time-dependent brane tension $\sigma = \sigma(t)$, as
we shall not be concerned with anisotropic branes. 
\section{minimal geometric deformation}
\label{MGD}
Solving the four-dimensional effective Einstein equations is not a straightforward task and,
already in the simple case of a spherically symmetric metric we shall consider here 
\be
\label{Sch}
ds^2=e^\nu\,dt^2-e^\lambda\,dr^2-r^2d\Omega^2
\ ,
\ee
only a few ``vacuum'' solutions are known analytically~\cite{wiseman,dejan,dadhich,page,cfabbri}.
When stellar systems are studied, the search for solutions becomes even more difficult,
mainly due to the presence of nonlinear terms in matter fields that arise from high-energy
corrections~\cite{maartens,GCGR}.
Nonetheless, two exact analytical solutions were found~\cite{ovalle2007}
by means of the MGD~\cite{jovalle2009}.
This approach has allowed in particular to generate physically acceptable interior 
solutions for stellar systems~\cite{jovalleBWstars}, to solve the tidally charged exterior
solution found in Ref.~\cite{dadhich} in terms of the ADM mass, and to study (micro)
black hole solutions~\cite{covalle1, covalle2}, as well as to elucidate the role
of exterior Weyl stresses from bulk gravitons on compact stellar
distributions~\cite{olps2013}.
\par
Let us start by reviewing the bases of the MGD approach, i.e.~the deformation
undergone by the radial metric component of the interior space-time associated with
a self-gravitating stellar system of radius $R$.
This radial metric component is deformed by bulk effects in such a way that,
when we demand to recover GR at low energies ($\sigma^{-1}\to 0$),
it must be written as
\begin{eqnarray}
\label{edlrwssg}
e^{-\lambda}
&=&
\mu(r)
+\underbrace{e^{-I}\int_0^r\frac{e^I}{\frac{\nu'}{2}+\frac{2}{x}}
\left[H(p,\rho,\nu)+\frac{k^2}{\sigma}\left(\rho^2+3\,\rho \,p\right)\right]
dx+\beta\, e^{-I}}_{\rm Geometric\ deformation}
\nonumber
\\
&\equiv&
\mu(r)+f(r)
\ ,
\end{eqnarray}
where
\begin{eqnarray}
\mu(r)
=
\begin{cases}
1-\strut\displaystyle\frac{\kappa_4^2}{r}\int_0^r x^2\,\rho\, dx
\equiv
1-\frac{2\,m(r)}{r}
\,,
&
\mbox{for}\quad r\,\leq\,R\,,
\\
\\
1-\strut\displaystyle\frac{2\,{M_0}}{r}
\ ,
&
\mbox{for}
\quad r>R
\ ,
\end{cases}
\end{eqnarray}
contains the usual GR mass function $m(r)$ for $r<R$, and ${M_0}$ for $r>R$,
whereas the function $H(p,\rho,\nu)$ encodes anisotropic effects due to bulk gravity
on the pressure $p$, matter density $\rho$ and the metric function $\nu$.
The function $\beta=\beta(\sigma)$ in Eq.~\eqref{edlrwssg} depends on the brane tension
$\sigma$ and on the mass $M_0$ of the self-gravitating system, and must be zero in the
GR limit.
For interior solutions, the condition $\beta(\sigma)=0$ must be imposed to avoid
singular solutions at the center $r=0$.
However, for a vacuum solution, or more properly, in the region $r>R$ where there is
a Weyl fluid filling the space-time surrounding the spherically symmetric stellar distribution,
the function $\beta$ is not necessarily zero, and there must be a geometric deformation
associated with the Schwarzschild solution.
\par
When a spherically symmetric GR vacuum solution is considered, the quantity $H=0$ and
the geometric deformation $f$ in vacuum ($p=\rho=0$), hereafter denoted $f=g^{*}(r)$,
will be consequently minimal and given by
\begin{equation}
\label{def}
g^*(r)
=\beta\,e^{-I}
\ .
\end{equation}
The radial metric component in Eq.~\eqref{edlrwssg} then becomes
\begin{eqnarray}
\label{g11vaccum}
e^{-\lambda}\,=\,
{1-\frac{2\,{M_0}}{r}}+\beta(\sigma)\,e^{-I}\, ,
\end{eqnarray}
where
\be
\label{I}
I
\equiv
\int\frac{\left(\nu''+\frac{{\nu'}^2}{2}+\frac{2\nu'}{r}+\frac{2}{r^2}\right)}
{\left(\frac{\nu'}{2}+\frac{2}{r}\right)}\,dr
\ .
\ee
\par
We then consider the general matching conditions between the generic interior MGD metric
\begin{equation}
\label{genint}
ds^2
=
e^{\nu^-(r)}-\frac{dr^2}{1-\frac{2m(r)}{r}+f^*(r)}-r^2d\,\Omega^2
\ ,
\end{equation}
characterizing the star interior $r<R$, where $f^*(r)$ is given by Eq.~\eqref{edlrwssg}
with $H=0$, and the most general exterior solution containing a Weyl fluid with
${\cal U}^+$, ${\cal P}^+$, and $p=\rho=0$ for $r>R$, which, according to the
expression in Eq.~\eqref{g11vaccum}, can be written as
\begin{equation}
\label{genericext}
ds^2
=
e^{\nu^+(r)} dt^2-\frac{dr^2}{1-\frac{2M}{r}+g^*(r)}-r^2d\Omega^2
\ ,
\end{equation}
where the mass $M$ in Eq.~\eqref{genericext} is in general a function of the brane tension $\sigma$.
Continuity of the first fundamental form at the star surface $\Sigma$ of radius $r=R$ when 
the metrics in Eq.~\eqref{genint} and \eqref{genericext} are considered, leads to
\begin{eqnarray}
\label{ffgeneric1}
\nu^-_R
&=&
\nu^+_R
\\
\nonumber
\\
\label{ffgeneric2}
\frac{2\, M}{R}
&=&
\frac{2\, M_0}{R}
+
\left(g^*_R-f^*_R\right)
\ ,
\end{eqnarray}
where $f_R^\pm\equiv f(r\to R^\pm)$ for any function.
Continuity of the second fundamental form on $\Sigma$ likewise gives~\cite{israel}
\be
\label{matching1}
\left[G_{\mu\nu}\,r^\nu\right]_{\Sigma}
=
0
\ ,
\ee
where $r_\mu$ is a unit radial vector and $[f]_{\Sigma}\equiv f(r\to R^+)-f(r\to R^-)$.
Using Eq.~\eqref{matching1} and Einstein field equations,
one finds $\left[T^{T}_{\mu\nu}\,r^\nu\right]_{\Sigma}=0$, which in our case reads
\be
\label{matching3}
\left[
p+\frac{1}{\sigma}\left(\frac{\rho^2}{2}+\rho\, p
+\frac{2}{k^4}\,\cal{U}\right)+\frac{4}{k^4}\,\frac{\cal{P}}{\sigma}
\right]_{\Sigma}
=0
\ .
\ee
Since we assume the distribution is only surrounded by a Weyl fluid
described by the functions ${\cal U}^+$ and ${\cal P}^+$,
$p=\rho=0$ for $r>R$, and this matching condition
takes the final form
\be
\label{matchingf}
p_R+\frac{1}{\sigma}\left(\frac{\rho_R^2}{2}+\rho_R\, p_R
+\frac{2}{k^4}\,{\cal U}_R^-\right)
+\frac{4}{k^4}\frac{{\cal P}_R^-}{\sigma}
=
\frac{2}{k^4}\frac{{\cal U}_R^+}{\sigma}+\frac{4}{k^4}\frac{{\cal P}_R^+}{\sigma}
\ ,
\ee
with $p_R\equiv p_R^-$
and $\rho_R\equiv \rho_R^-$.
\par
The limit $\sigma^{-1}\rightarrow 0$ in the second fundamental form in
Eq.~\eqref{matchingf} leads to the well-known GR matching condition $p_R =0$
at the star surface.
The expressions given by Eqs.~\eqref{ffgeneric1}, \eqref{ffgeneric2} and~\eqref{matchingf}
are the necessary and sufficient conditions for the matching of the interior MGD metric
to a spherically symmetric ``vacuum'' filled by a BW Weyl fluid~\cite{germ,gergely2006}. 
\section{Black strings and variable tension brane}
\label{BS}
In this Section, we proceed to apply the above formalism to the case of a black string.
We shall first derive the brane geometry from the MGD approach applied to the standard 
Schwarzschild metric, in order to obtain the relevant projections of the bulk
Weyl tensor that enter the bulk metric~\eqref{tay}.
Subsequently, we shall study the particular case of a phenomenological E\"otv\"os brane,
with a variable tension that depends on the time exponentially.
\subsection{Brane geometry of a black string}
\label{IVA}
Let us now find the explicit MGD function $g^*(r)$ produced by the Schwarzschild
solution
\be
\label{Schw}
e^{\nu_S}=e^{-\lambda_S}
=
1-\frac{2\,{M}}{r}
\ ,
\ee
where we recall $M$ is a function of the brane tension $\sigma$.
Using Eq.~\eqref{Schw} in Eq.~\eqref{def}, we obtain
\be
\label{defS}
g^*(r)=
-\frac{2\,\beta(\sigma)}{r}\,\frac{1-\frac{2M}{r}}{r-\frac{3M}{2}}
\ ,
\end{equation}
and the deformed exterior metric components read
\ba
\label{schoo}
e^{\nu}
&=&
1-\frac{2\,{M}}{r}
\ ,
\\
e^{-\lambda}
&=&
\left(1-\frac{2\,{M}}{r}\right)
\left[1-\frac{\beta(\sigma)}{r-\frac{3\,{M}}{2}}\right]
\ ,
\label{sch11}
\ea
which match the vacuum solution found in Ref.~\cite{mage} in the particular case
when $\beta(\sigma)=-\frac{C_0}{\sigma}$, for $C_0$ a positive constant. 
Below we shall show a general expression for the function $\beta$,
which depends on the interior structure of the self-gravitating system surrounded
by the geometry~\eqref{schoo}.
The Weyl fluid associated with the solution in Eqs.~\eqref{schoo}, \eqref{sch11}
is then described by the functions (see, e.g.~Ref.~\cite{olps2013})
\ba
\label{pp2}
\frac{1}{k^2}\,\frac{{\cal P}^+}{\sigma}
&=&
-\frac{\left(1-\frac{4\,{M}}{3\,r}\right)}{9\left(1-\frac{3\,{M}}{2\,r}\right)^2}\,
\frac{\beta}{r^3},\qquad {\rm and}\qquad\quad
\frac{1}{k^2}\,\frac{{\cal U}^+}{\sigma}
=
\frac{{M}}{12\left(1-\frac{3\,{M}}{2\,r}\right)^2}\frac{\beta}{r^4}
\ .
\ea
The function $\beta=\beta(\sigma)$ must now be specified by considering the
deformed Schwarzschild solution~\eqref{schoo} and \eqref{sch11} in the matching
conditions~\eqref{ffgeneric1}, \eqref{ffgeneric2} and~\eqref{matchingf}.
The first fundamental form leads to the expressions given by  Eqs.~\eqref{ffgeneric1}
and \eqref{ffgeneric2}, where Eqs.~\eqref{ffgeneric1} becomes
\be
\label{minmatch1}
e^{\nu^-_R}
=1-\frac{2{M}}{R}
\ ,
\ee
whereas the second fundamental form in Eq.~\eqref{matchingf} gives
\be
\label{sfgeneric}
p_R+\frac{f^*_R}{k^2}\left(\frac{\nu'_R}{R}+\frac{1}{R^2}\right)
=
-\frac{g^*_R}{R^2}
\ , 
\ee
Note that if $M$ were the GR mass $M_0$ of Eq.~\eqref{Schw}),
one would have $g_{R}^*=f_{R}^*$, which represents an unphysical condition, 
according to the expression in Eq.~\eqref{sfgeneric}).
Eqs.~\eqref{ffgeneric2}, \eqref{minmatch1} and~\eqref{sfgeneric} are the
necessary and sufficient conditions for matching the two minimally deformed
metrics given in Eq.~\eqref{genint} and in Eqs.~\eqref{schoo} and \eqref{sch11}.
The matching condition~\eqref{sfgeneric} shows that the exterior geometric deformation
$g^*(r)$ at the star surface, i.e.~$g^*_R$, is {\em always negative\/}.
Therefore, the deformed horizon radius $r_h=2\,{M}$ will always be smaller than the
Schwarzschild radius $r_H=2\,M_0$, as can clearly be seen from Eq.~\eqref{ffgeneric2}.
This general result clearly shows that five-dimensional effects weaken the 
strength of the gravitational field produced by the self-gravitating stellar system.
\par
Finally, when the explicit geometric deformation~\eqref{defS} is considered in the
matching condition~\eqref{sfgeneric}, the function $\beta=\beta(\sigma)$ becomes
\begin{equation}
\label{beta}
\beta(\sigma)
=
R^3\left(\frac{1-\frac{3\,{M}}{2\,R}}{1-\frac{2\,{M}}{R}}\right)
\left[\left(\frac{\nu'_R}{R}+\frac{1}{R^2}\right)\frac{f^*_R}{8\pi}+p_R\right]
\ ,
\end{equation}
showing thus that $\beta$ is always positive and (interior) model-dependent.
For instance, we can find $\beta(\sigma)$ by considering the exact interior BW solution
found in Ref.~\cite{ovalle2007}, where the geometric deformation is given by
\ba
\label{regularmass}
f^*(r)
&=&
\frac{1}{\sigma}\,\frac{4C(\tau(r))}{49\pi}
\left[\frac{240+589Cr^2-25C^2r^4-41C^3r^6-3C^4r^8}{3(1+Cr^2)^2}
-\frac{80\,{\rm \arctan}(\sqrt{C}r)}{\sqrt{C}r}\right]
\ ,
\ea
with $C$ a constant given by $CR^2=\frac{\sqrt{57}-7}{2}\equiv\alpha$,
and the functions $(\tau(r))^{-1}\equiv (1+Cr^2)^3(1+3Cr^2)$
and $\nu'=\frac{8Cr}{1+Cr^2}$. 
Now, by using the explicit form of $f(R)$ we get
\begin{eqnarray}
\label{betafinal2}
\beta
=
\frac{\alpha\,\tau(R)}{98\pi^2\sigma}\,
\frac{1-\frac{3\,{M_0}}{2\,R}}{1-\frac{2\,{M_0}}{R}}
\left[1+9\,\alpha\right]
\left[\frac{240+589\,\alpha-25\,\alpha^2-41\,\alpha^3-3\,\alpha^4}{3(1+\alpha)^2}
-\frac{80\,{\rm \arctan}(\sqrt{\alpha})}{\sqrt{\alpha}}\right]
\ ,
\end{eqnarray}
or
\begin{equation}
\label{betafinal3}
\beta(\sigma)
\propto
\frac{1}{2\,\sigma\,R}
\frac{2\,R-{3\,{M_0}}}{R-{2\,{M_0}}}
\equiv
\frac{C_0}{\sigma}
\ .
\end{equation}
Note that we are using $M_0$ and not $M$, because $M=M_0+{\cal O}(\sigma^{-1})$,
so that the exact $\beta=\beta(\sigma)$ will be given by the expression above
plus terms of order $\sigma^{-2}$. 
\par
Now, as the area of the five-dimensional horizon is determined by
$g_{\theta\theta}(x^\alpha,y)$, we need to find ${\cal E}_{\theta\theta}$
determined by the deformation $f=g^*(r)$ shown in Eq.~\eqref{defS}.
Upon using Eq.~\eqref{Sch}, we readily find 
\be
\label{E22}
{\cal E}_{\theta\theta}
=
-R_{\theta\theta}
=
\frac{r}{2}\,e^{-\lambda}
\left(\lambda'-\nu'\right)
+1-e^{-\lambda} 
=
\frac{\beta(\sigma)}{2}\frac{\left(1-\frac{M}{r}\right)}{\left(1-\frac{3\, {M}}{2\,r}\right)^2}
\ ,
\ee
This is the component of the projection of the Weyl tensor on the brane which we
need in order to determine the horizon area in the bulk, where the function
$\beta=\beta(\sigma)$ can be specified once we choose a specific interior BW solution
to evaluate the expression in Eq. \eqref{beta} and a model of the brane tension.
\subsection{Time-dependent horizon area}
As the main object of interest to us here is the black string horizon along the extra dimension, 
we shall focus on the term $g_{\theta\theta}$ which yields the area of the horizon,
and evaluate the corresponding expansion~\eqref{tay}.
Clearly, a time-dependent brane tension will modify the black string Schwarzschild
background. 
Since the complete solution is extremely difficult to compute, we shall take
a more effective approach, and study how the horizon area changes with the brane
tension in the Taylor expansion~\eqref{tay}.
As we shall show, even in this approximate description, very interesting results
can be obtained.
\par
Upon inserting the relevant expressions~\eqref{schoo}, \eqref{sch11} and~\eqref{E22}
derived from the MGD of the Schwarzschild metric, the Taylor expansion of the term
$g_{\theta\theta}(x^\alpha,y)=g_{\theta\theta}(r,y)$ in Eq.~\eqref{tay} reads
\ba
g_{\theta\theta}(t,r,y)
&=&
r^2-\frac{r^2}{3}\,\kappa_5^2\,\sigma\,|y|
+\left(\frac{\kappa_5^4\sigma^2}{36}- \frac{\Lambda_5}{6}\right)r^2 y^2
\nonumber
\\
&&
-\left[\frac{193}{216}\sigma^3\kappa_5^6 
+\frac{5}{18}\Lambda_5\kappa_5^2\sigma
+\frac{\kappa_5^2\,\beta(\sigma)}{12\,r}\,\frac{\left(1-\frac{M}{r}\right)}{\left(1-\frac{3M}{2\,r}\right)^2}
\right]
\frac{r^2|y|^3}{3!}
\nonumber
\\
&&
+
\left\{
\frac{\Lambda_5}{18}
\left(\Lambda_5-\frac{3\beta(\sigma)}{2\,r}\frac{\left(1-\frac{M}{r}\right)}{\left(1-\frac{3M}{2\,r}\right)^2}
- \frac{\sigma^2\kappa_5^4}{6}
+\frac{7\,\sigma^4\kappa_5^8}{324}\right)
\right.
\nonumber
\\
&&
\left.
\quad
+
\frac{\left(1-\frac{2M}{r}\right)}{\left(1-\frac{3M}{2\,r}\right)}
{r^4}\,M \left(2 \beta + 3 M - 2 r\right)
\left[54 M^3 r + 9 M^2 \left(1 + 4 \beta r - 10 r^2\right)
\right.
\right.
\nonumber
\\
&&
\left.
\left.
\phantom{AB}
+2 r^2 \left(2 + 7 \beta r - 4 r^2\right)
+ 12 M r \left(4 r ^2-\beta\right) r-1
\right]
+ \left(3 M - 2 r\right)^2
\right.
\nonumber
\\
&&
\quad
\left.
-\left(\frac{5\Lambda_5}{6}+\frac{83\sigma^2\kappa_5^4}{216}\right)
\frac{\beta(\sigma)}{2\,r}\frac{(1-\frac{M}{r})}{(1-\frac{3M}{2\,r})^2}
\right\}
\frac{r^2 y^4}{4!}
+\cdots
\ ,
\label{expp}
\ea
where the time dependence enters via $\sigma=\sigma(t)$.
In particular, the extra terms of order $|y|^3$ in Eq.~\eqref{addtruey3}
can be reduced to
\be
g_{\theta\theta}^{(3)\,{}^{\rm variable}}(t,r)
=
-\frac{2}{3}\kappa_5^2\sigma''r^2
\ ,
\label{y33}
\ee
where, from now on, $\sigma'\equiv d\sigma/d t$ and so on.
Furthermore, at order $y^4$ we have the terms in Eq.~\eqref{addtruey4}
are given by
\be
g_{\theta\theta}^{(4)\,{}^{\rm variable}}(t,r)
=
\kappa_5^2\,r^2
\left[\frac{\kappa_5^2\sigma^2}{9}
-\frac{\sigma^{\prime\prime\prime\prime}}{3} 
+\frac{\sigma^{\prime\prime}}{9}
\left(\frac{197}{72}\sigma^2\kappa_5^2 
- \Lambda_5-6\,\beta(\sigma)\,\frac{r-2M}{2\,r-3M}\right)
\right]
\ .
\label{termse}
\ee
All the expressions obtained are so far the most general.
In order to better understand the physical implications of a variable tension 
on the event horizon along the extra dimension,
let us apply our analysis to a specific physically motivated case.
\subsection{Black string with minimal geometric deformation in a brane variable tension}
A variable tension on the brane can be established in general by two approaches.
On the one hand, one can consider the brane tension as a scalar field
in the Lagrangian, as is widely assumed in the context of string theory 
and supersymmetric branes~\cite{CORDAS,SUSYBRANES}.
Otherwise, as we shall do in the analysis hereupon, the brane tension can be
understood as an intrinsic property of the brane~\cite{GERGELY2008,GERGELY2009,PRD,EPJC}. 
\par
In the BW scenario, the functional form of the variable brane tension is an open issue.
However, taking into account the huge variation of the universe temperature during its
cosmological evolution, it is indeed plausible to implement the brane tension as a 
function of the cosmological time.
Although it lacks a complete scenario, the phenomenologically interesting case 
of E\"otv\"os fluid membranes~\cite{Eotvos} can be useful to extract physical results.
In this context, the cosmological evolution of a perfect fluid imposes cosmological
symmetries, and the brane tension $\sigma$, along with the constants $\kappa_4$ and
$\Lambda_4$, become scale-factor (or cosmological time) dependent. 
The phenomenological E\"otv\"os law asserts that the fluid membrane tension
depends on the temperature as
\be
\sigma
=\chi
\left(T_{c}-T\right)
\ ,
\label{TERM0}
\ee 
where $\chi$ is a constant and $T_{c}$ represents a critical temperature equal to
the highest temperature for which the membrane exists.
By imposing the continuity equation, in such a way that the temperature dependence
of the brane tension is balanced by the energy exchange between the brane and the bulk,
it implies that the brane is formed in a very hot early universe when $\sigma \simeq 0$,
and initially $\kappa_4$ and $\sigma$ as well are small, strengthening BW effects.
If there are no stresses in the bulk, and without taking into account the cosmological
constant, the bulk is isolated from the brane, with no exchange of
energy-momentum~\cite{maartens}.  
Thus the thermodynamical expression $dQ=dE+p\,dV=0$ holds for the brane.
Furthermore, for photons of the CMB in a volume $V$, we can use
$E=E_{\gamma}=\sigma\, T^{4}\,V$ and $p=\frac{E}{3}=\frac{\sigma T^4}{3}\,V$.
It is then easy to verify that $ \frac{1}{T}\frac{dT}{dt}=-\frac{1}{3V}\frac{dV}{dT}$.
By relating the volume to the Friedmann-Robertson-Walker scale factor,
one obtains $T \propto \frac{1}{a(t)}$~\cite{bs1,GERGELY2009}.
By considering the similarity with E\"otv\"os membranes, one then infers that
$\sigma=\sigma_{0}\left(1-\frac{a_{0}}{a}\right)$, where $\sigma_{0}$ is a scalar
associated to the four-dimensional coupling constants~\cite{GERGELY2009},
and $a_{0}$ is the minimum scale factor such that the brane does not exist
(meaning that for smaller $a$, the brane tension would be negative).
In any case, in what follows, it suffices to consider
\be
\sigma(t)=1-\frac{1}{a(t)}
\ .
\ee
We can also note that a similar behaviour is obtained from SUSY in
inflationary cosmology. 
In fact, as shown in Ref.~\cite{bs1}, one obtains
\be
{\Lambda_{4}}
\propto
\left(1-\frac{1}{a(t)}\right)^2\,\label{la4}
\ ,
\ee
which means that the cosmological constant takes negative values before it becomes 
positive, as the universe expands.
\par
To be more specific, a de~Sitter brane profile is taken into account by
setting $a(t)\propto e^{\gamma\, t}$, with positive $\gamma$~\cite{RE41},
so that
\be
\sigma(t)
=
1-e^{-\gamma\, t}
\ .
\label{ULTI}
\ee
The phenomenological viability of this model was analysed in Ref.~\cite{bs1}. 
Further, we shall set $\Lambda_5 = 1=\kappa_5$ from here on.
As the brane tension has the lower bound
$\sigma \sim 4.39 \times 10^8\,$MeV$^4$~\cite{maartens,mage,1jea},
we shall normalize it accordingly in the analysis below.
\par
\begin{figure}[h]
\begin{center}\includegraphics[width=3.2in]{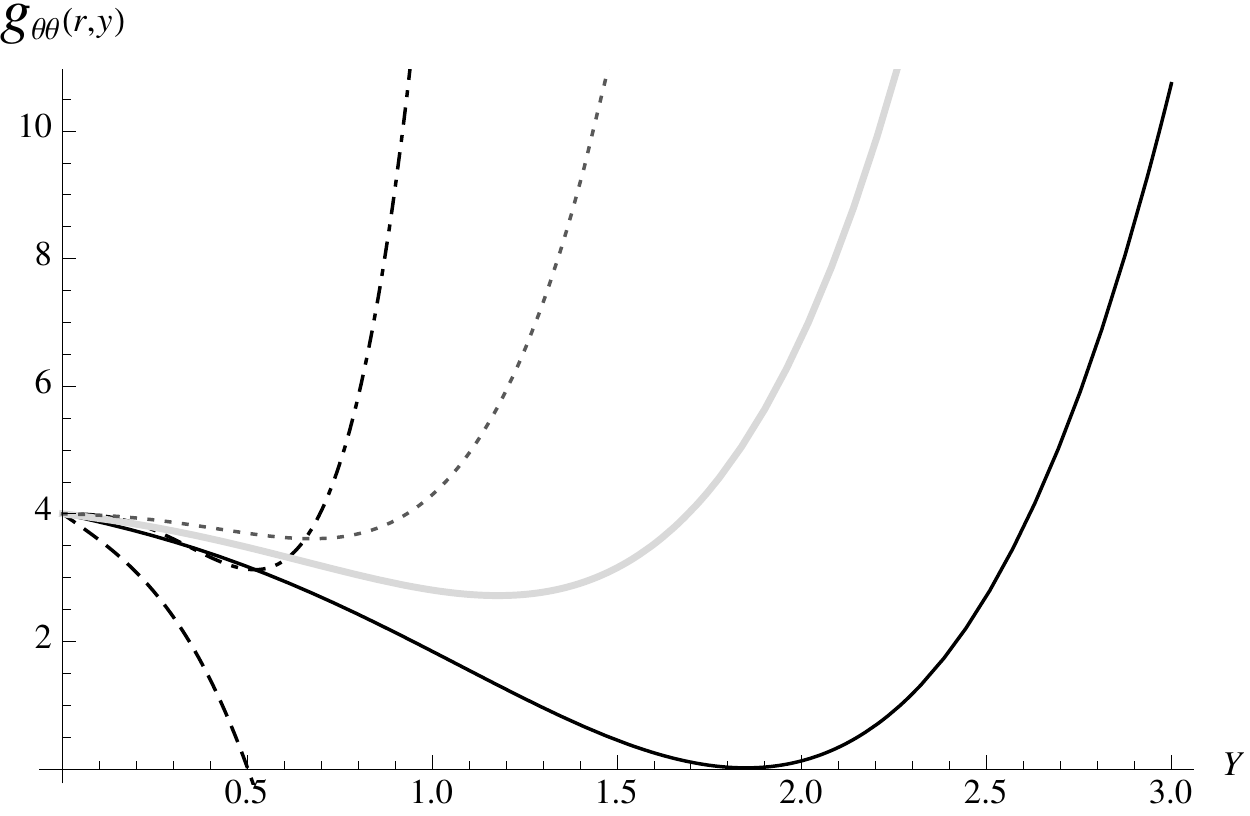}  
\caption{Metric element $g_{\theta\theta}(r=2\,M,y)$ 
 along the extra dimension
for $\sigma = 2$ (dashed line), $\sigma = 0.92$ (solid black line), 
$\sigma = 0.5$ (thick gray line), $\sigma = 0.1$ (dotted line),
$\sigma = 0.05$ (dash-dotted line).
Black hole mass $M=1$ and constant $C_0=1$.
 \label{fig1}}
\end{center}
\end{figure}
The first result we present is the plot of the area of slices of constant $y$ of the
black string horizon $r_H=2\,M$ for different constant values of the brane tension
$\sigma$ (see Fig.~\ref{fig1}).
It is worth noticing that when the brane tension reaches the value $\sigma \approx  0.92$,
the black string warped horizon changes profile:
for $\sigma \gtrsim 0.92$ the horizon area is always positive, whereas for $\sigma \lesssim 0.92$,
there is always a point of coordinate $y_c$ along the extra dimension where the horizon meets
the axis of axial symmetry $g_{\theta\theta}(r=r_H,y_c)=0$. 
For instance, when $\sigma = 0.05$, one finds $y_c \simeq 0.51$ in Fig.~\ref{fig1}.
\par
\begin{figure}[h]
\begin{center}\includegraphics[width=3.2in]{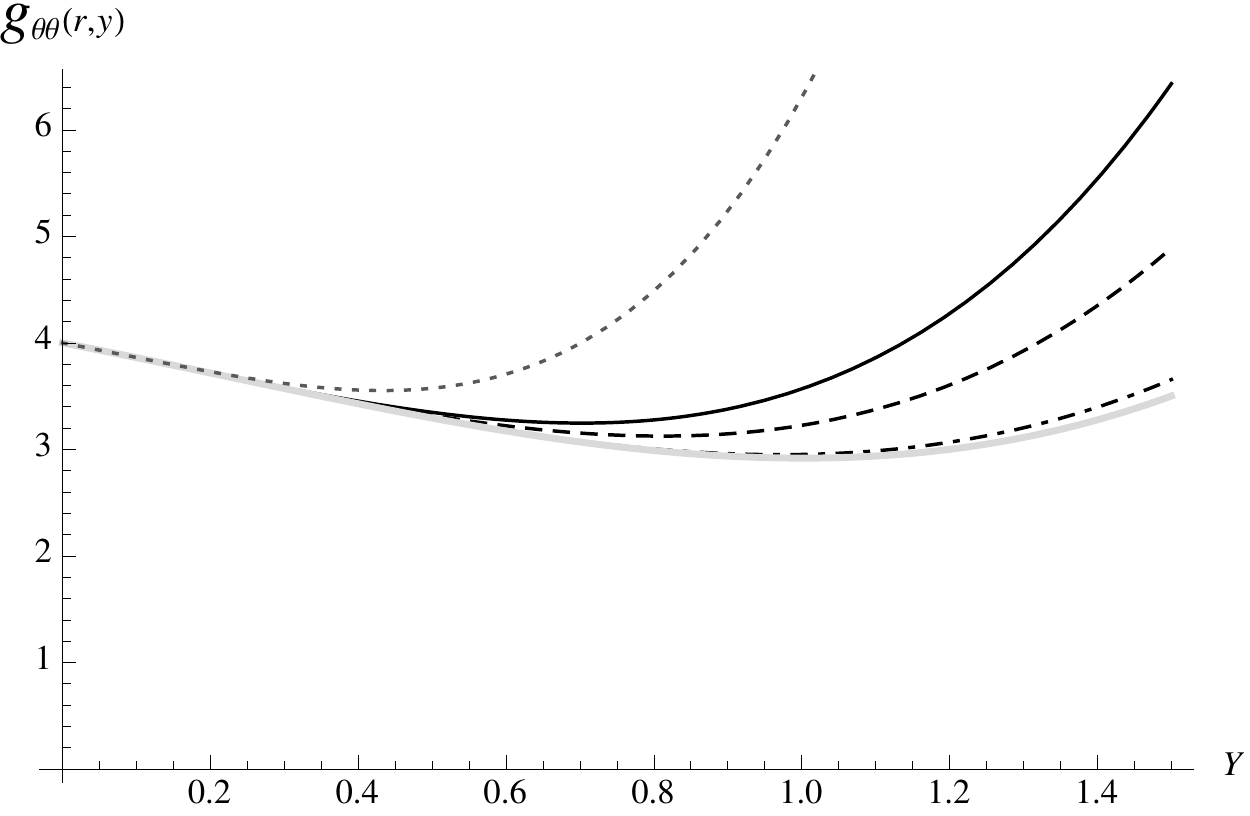}
\caption{Metric element $g_{\theta\theta}(r=2\,M,y)$ along the extra dimension for
$C_0 = 5$ (black dotted line), $C_0 = 1$ (black solid line),
$C_0 = 0.5$ (dashed line), $C_0 = 0.1$ (dot-dashed line), $C_0 = 0.05$
(thick gray line).
Black hole mass $M=1$ and brane tension $\sigma=1$.
\label{fig2}}
\end{center}
\end{figure}
Next, in Fig.~\ref{fig2} one can see how the black string warped horizon varies along the
extra dimension as a function of $\beta(\sigma)=C_0/\sigma$ from the MGD described
in Section~\ref{IVA}. 
As the brane tension $\sigma$ is assumed heretofore constant, we have rescaled
the parameter $C_0$ accordingly.
\par
\begin{figure}[h]
\begin{center}
\includegraphics[width=3.2in]{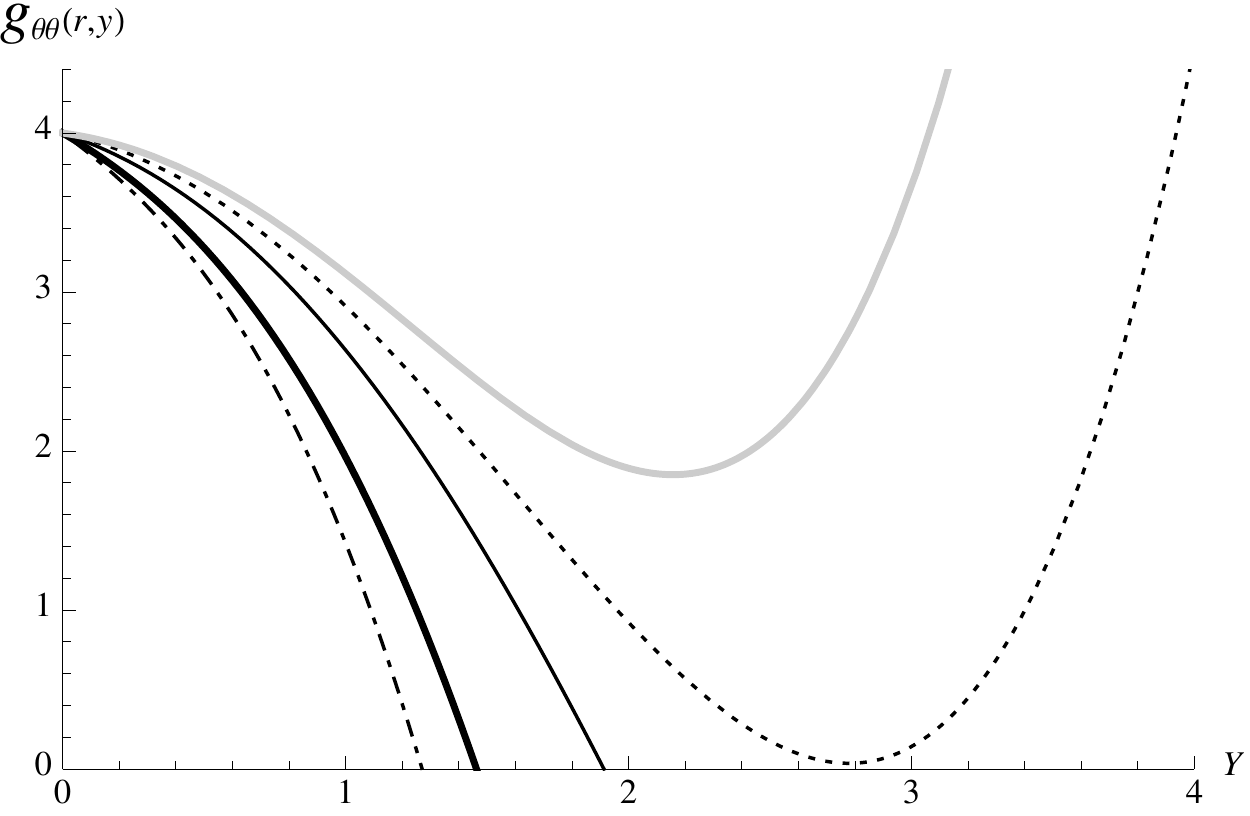}
\caption{
Metric element $g_{\theta\theta}(t,r=2M,y)$ along the extra dimension for $C_0 = 0.1$
and $\gamma\, t = 0.1$ (thick gray line), $\gamma\, t = 0.176$ (dotted line - this is the
point where the horizon area goes to zero), $\gamma\, t= 0.75$ (thick black line),
$\gamma\,t=1$ (black solid line), $\gamma\,t=2$ (dot-dashed line).
\label{fig3}}
\end{center}
\end{figure}
\begin{figure}[h]
\begin{center}\includegraphics[width=4in]{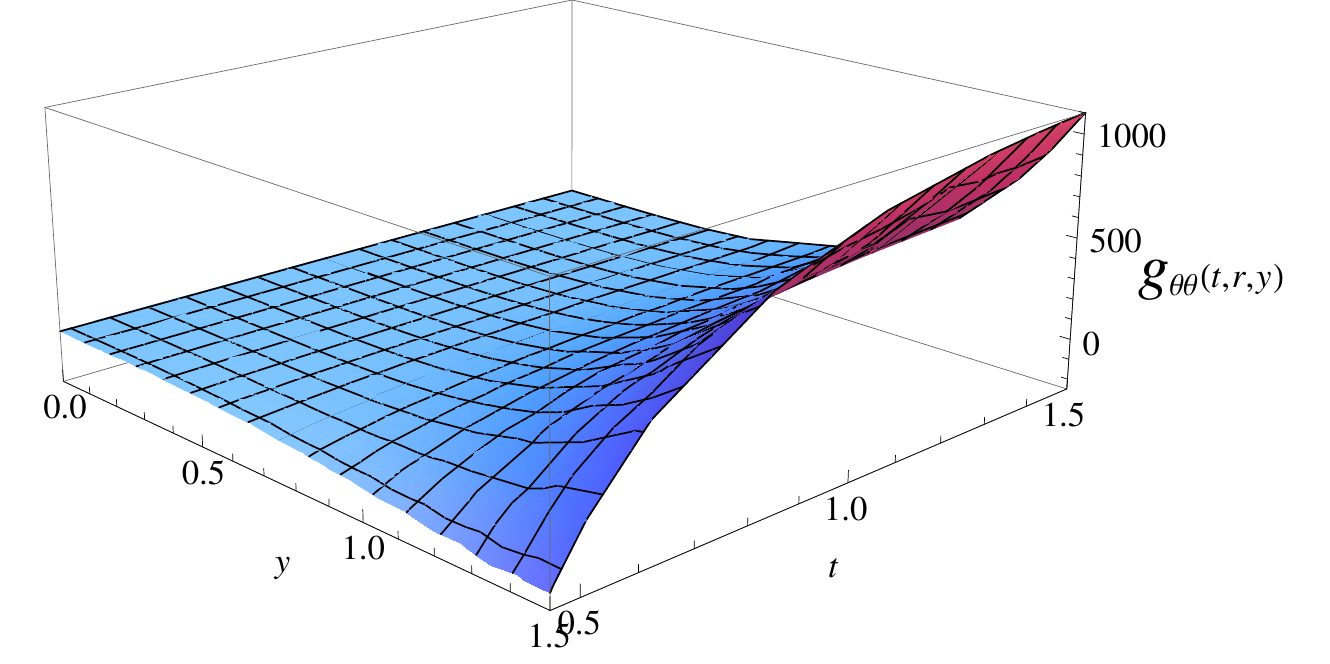}
\caption{Metric element $g_{\theta\theta}(t,r=2M,y)$ along 
the extra dimension as a function of time, for $C_0 = 10$.
\label{fig4}}
\end{center}
\end{figure}
Of course, there is no time dependence in the cases shown in Figs.~\ref{fig1} and~\ref{fig2}.
The plot in Fig.~\ref{fig3} instead displays the black string warped horizon given by the bulk
metric~\eqref{expp} on the horizon $r=2\,M$, with the additional terms~\eqref{y33}
and \eqref{termse} due to the variable brane tension given by the expression~\eqref{ULTI}.
One can also see from Fig.~\ref{fig4} that, for values $\gamma\,t\gtrsim 0.52$,
the horizon area increases monotonically along the extra dimension.
For $\gamma\,t\lesssim 0.52$ there exists a point $y_c$ along the extra dimension
beyond which the black string ceases to exist.  
We shall further discuss this point below.
\par
\begin{figure}[h]
\begin{center}
\includegraphics[width=3in]{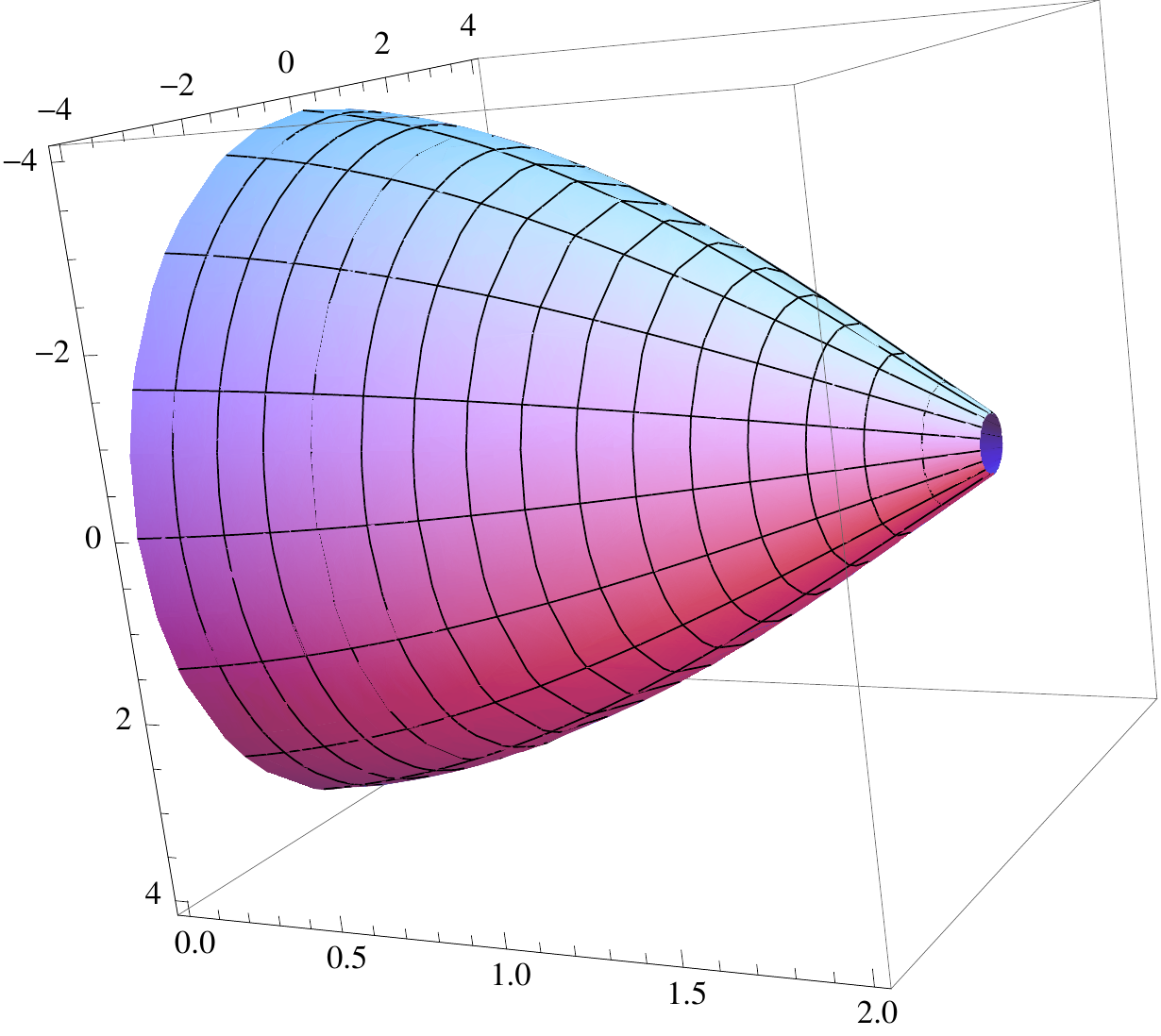}
\caption{Warped horizon along the extra dimension, for $t=0.3$.
Here $C_0 = 0.1$ and $\gamma\, t = 0.25$, for a variable brane tension. 
\label{fig5}}
\end{center}
\end{figure}
In Fig.~\ref{fig5}, the variable brane tension is considered for $\gamma\,t = 0.25$.
The coordinate $y$ along the extra dimension beyond which
the black string ceases to exist is $y_c \simeq 2.1$.
Note that for $y>y_c$, the metric element $g_{\theta\theta}(t,r=2M,y>y_c)<0$,
and such negative values are also displayed for clarity but have no physical meaning.
Moreover, the four-dimensional Kretschmann scalar
$ R_{\mu\nu\rho\sigma}\,R^{\mu\nu\rho\sigma}$ diverges for $r = 0$.
This can be seen from the Gauss equation which relates the four- and five-dimensional
Riemann curvature tensors according to
${}^{(5)}R^\mu _ {\;\; \nu\rho\sigma} = {}^{(4)}R^\mu_{\;\; \nu\rho\sigma}
 -K^\mu _ {\;\; \rho}K_{\nu\sigma} + K^\mu _ {\;\; \sigma}K_{\nu\rho}$.  
Now, Eq.~\eqref{curv} in this specific case reads 
$K_{\mu\nu}=-\frac{\kappa_5^2}{6} \sigma g_{\mu\nu}$ and, by inserting it in the
Gauss equation for the five-dimensional Kretschmann scalar 
$ {}^{(5)}K_0 =  {}^{(5)}R_{\mu\nu\rho\sigma} {}^{(5)}R^{\mu\nu\rho\sigma}$,
one can see that terms involving the extrinsic curvature do not  cancel the
divergence provided by the four-dimensional  Kretschmann scalar in $ {}^{(5)}K_0$.
For such cases, the Kretschmann scalar diverges at $r \to 0$, and
also $K_0$ diverges at $y = y_c$, when  $r=0$, characterising indeed a singularity
at $y=y_c$.
\par
\begin{figure}[H]
\begin{center}
\includegraphics[width=2in]{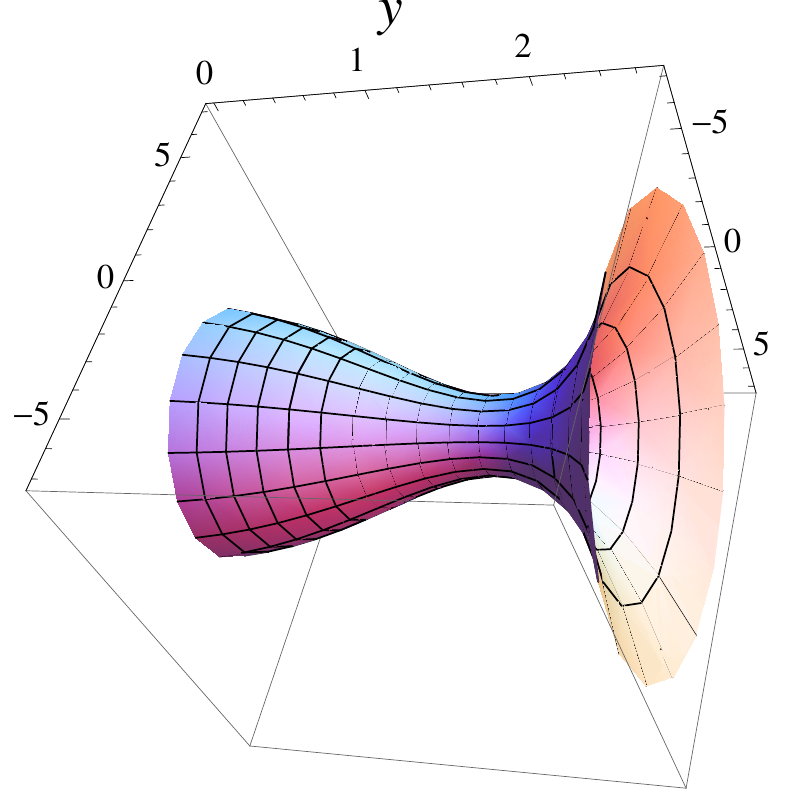}
\includegraphics[width=2in]{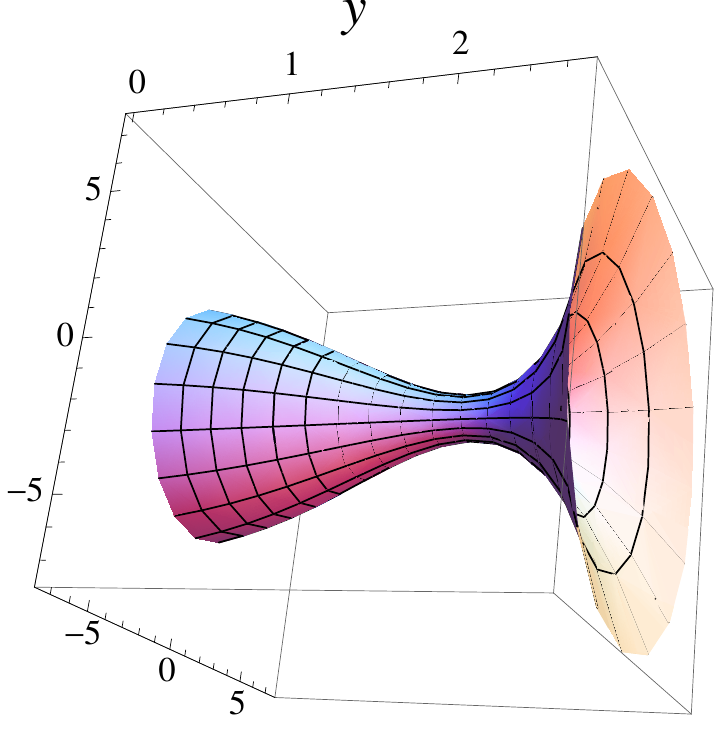}
\includegraphics[width=2.1in]{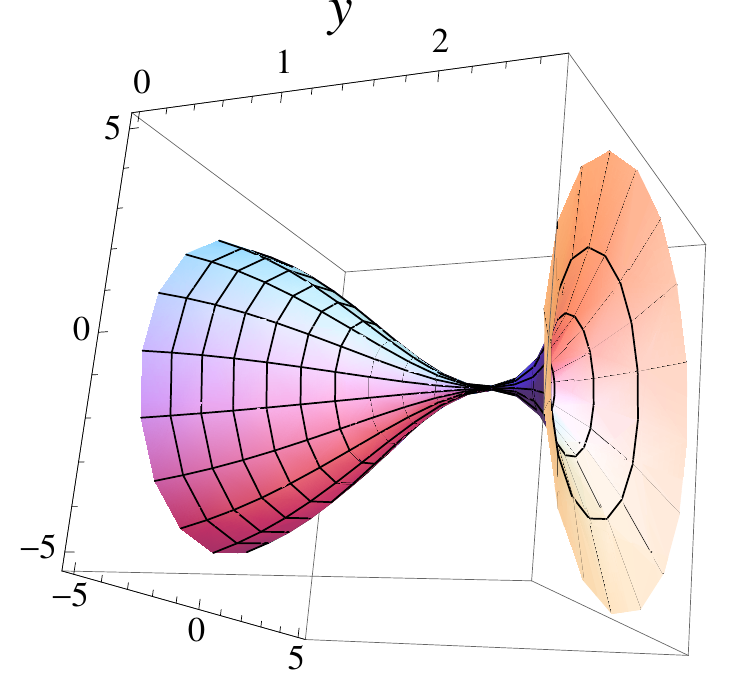}
\caption{Warped horizon along the extra dimension for $\gamma\,t=0.5$, $\gamma\,t=1.25$
and $\gamma\,t=3$ (left to right). Here $C_0 = 1$.
\label{fig6}}
\end{center}
\end{figure}
The plots in Fig.~\ref{fig6} display the black string warped horizon for increasing
values of the time variable.
As the time passes, the horizon pinches at a critical point along the extra dimension.  
In the context of the MGD of a black hole on the variable tension brane,
the black string presents a throat along the extra dimension, with area that tends to zero
as time runs to infinity. 
As it is going to be discussed in the next Section, our results completely support the ones
obtained in Ref.~\cite{Horowitz:2001cz,kol}, in the context of the MGD.
The area theorem prevents indeed the throat from shrinking to zero size in finite horizon time.
\section{Concluding Remarks}
It is well-known that BW models lead to cosmological evolutions whose background dynamics
is completely understood, and can further reproduce general relativistic results with suitable
restrictions on the BW parameters.
In this paper, we focused our attention on branes whose variable tension only depends on
the time, and obtained the Taylor expansion of the metric in the Gaussian coordinate
along the extra dimension.
\par
This Taylor expansion makes it possible to write the bulk metric in terms of the brane metric,
and in a form that shows a time-dependence already in terms of the second order.
It is however only including terms of (at least) third order that one can display the effects
of a variable tension on the shape of the horizon in the bulk, since it is only from such terms 
that the covariant derivatives of the variable tension appear in the metric expansion along
the extra dimension, and lead to sizeable corrections.
Furthermore, as in some cosmological epochs the value of the tension could have been
very small, and can thus be largely modified, it is very important to consider such terms,
which contribute to a realistic description of black strings.
The additional terms in the expansion we investigated indeed alter the black string
warped horizon, when the brane tension varies in a BW model based upon the
E\"otv\"os law. 
\par
We also showed that the MGD of a Schwarzschild black hole on the variable tension
brane corresponds to a black string with a throat along the extra dimension, whose area
tends to zero as time runs to infinity.
\par
One could finally note that Eq.~\eqref{la4} implies that, for small values of the scale factor,
the effective four-dimensional cosmological constant $\Lambda_4 < 0$, and it
contributes like an attracting cosmic component.
As the scale factor increases, one reaches $\Lambda_4 > 0$, which generates a
dark energy-type repulsion. 
In addition, BW models with varying brane tension might allow for energy exchange
and other types of evolution that concretely lead to interesting new physics,
supporting and generalizing some results in the literature, as those in
Ref.~\cite{Gergely:2006hd,Mak:2004hv,gharko} for instance.
In particular, BW models can replace dark matter with geometric effects,
and we plan to investigate BW stars and the gravitational collapse on the
brane~\cite{germ,jovalleBWstars, ovalle2007,casadio1,Bruni:2001fd, Gergely:2006xr}
by means of generalizations of spherically symmetric BW solutions
to the case with variable brane tension. 
\section*{Acknowledgments}
R.~Casadio is supported in part by the European Cooperation in Science and Technology (COST)
action MP0905 ``Black Holes in a Violent  Universe".
R.~da~Rocha is grateful to CNPq grant 303027/2012-6 for partial financial support.

\end{document}